\documentclass[prd,preprintnumbers,12pt]{revtex4}
\usepackage{epsfig}
\usepackage{graphicx}
\usepackage{amsmath}
\usepackage{amssymb}

\newcommand{\be}{\begin{equation}}

\newcommand{\ee}{\end{equation}}
\newcommand{\bea}{\begin{eqnarray}}
\newcommand{\eea}{\end{eqnarray}}

\newcommand{\nn}{\nonumber}

\begin{document}
\title{Effective fermion couplings  in warped 5D Higgsless theories}

\author{J. Bechi, R. Casalbuoni, S. De Curtis and D. Dominici}
\affiliation{Department of Physics, University of Florence, and
INFN, Florence, Italy}

\date{\today}

\begin{abstract}
\noindent
We consider a five dimensional $SU(2)$  gauge theory with fermions in the bulk and with additional $SU(2)$ and $U(1)$ kinetic terms on the branes. The  electroweak breaking is obtained by  boundary conditions. After deconstruction,
fermions in the bulk are eliminated by using  their equations of motion.
In this way Standard Model  fermion mass terms   and   direct couplings to the internal gauge bosons of the moose are generated.
The presence of these new couplings gives  a new contribution to the $\epsilon_3$ parameter in addition to the gauge boson term. This allows
the possibility of a cancellation between the two contributions, which can be local
(site by site) or global.
Going back to the continuum, we show that the implementation of local cancellation in any generic  warped metric
leaves massless fermions. This is due to the presence of one horizon on
the infrared  brane. However we
 can  require a global cancellation of the new physics contributions to the $\epsilon_3$
parameter. This fixes relations among the warp factor and the parameters of the fermion
and gauge sectors.

\end{abstract}
\pacs{xxxxxxx} \maketitle
\newcommand{\f}[2]{\frac{#1}{#2}}
\def\to{\rightarrow}
\def\ptl{\partial}
\def\beq{\begin{equation}}
\def\eeq{\end{equation}}
\def\bea{\begin{eqnarray}}
\def\eea{\end{eqnarray}}
\def\nn{\nonumber}
\def\half{{1\over 2}}
\def\rhalf{{1\over \sqrt 2}}
\def\calo{{\cal O}}
\def\call{{\cal L}}
\def\calm{{\cal M}}
\def\del{\delta}
\def\eps{\epsilon}
\def\lam{\lambda}

\def\anti{\bar}
\def\delfac{\sqrt{{2(\del-1)\over 3(\del+2)}}}
\def\heff{h'}
\def\square{\boxxit{0.4pt}{\fillboxx{7pt}{7pt}}\hspace*{1pt}}
    \def\boxxit#1#2{\vbox{\hrule height #1 \hbox {\vrule width #1
             \vbox{#2}\vrule width #1 }\hrule height #1 } }
    \def\fillboxx#1#2{\hbox to #1{\vbox to #2{\vfil}\hfil}   }

\def\braket#1#2{\langle #1| #2\rangle}
\def\gev{~{\rm GeV}}
\def\gam{\gamma}
\def\sn{s_{\vec n}}
\def\sm{s_{\vec m}}
\def\mm{m_{\vec m}}
\def\mn{m_{\vec n}}
\def\mh{m_h}
\def\sumn{\sum_{\vec n>0}}
\def\summ{\sum_{\vec m>0}}
\def\vl{\vec l}
\def\vk{\vec k}
\def\ml{m_{\vl}}
\def\mk{m_{\vk}}
\def\gp{g'}
\def\gt{\tilde g}
\def\hw{{\hat W}}
\def\hz{{\hat Z}}
\def\ha{{\hat A}}

\def\yy{{\cal Y}_\mu}
\def\yyt{{\tilde{\cal Y}}_\mu}
\def\lq{\left [}
\def\rq{\right ]}
\def\dmu{\partial_\mu}
\def\dnu{\partial_\nu}
\def\dmus{\partial^\mu}
\def\dnus{\partial^\nu}
\def\gp{g'}
\def\gpt{\tilde g'}
\def\gptt{\tilde g^{\prime 2}}
\def\gs{g''}
\def\ggs{\frac{g}{\gs}}
\def\tr{{\rm {tr}}}
\def\V{{\bf{V}}}
\def\W{{\bf{W}}}
\def\Wt{\tilde{ {W}}}
\def\Y{{\bf{Y}}}
\def\Yt{\tilde{ {Y}}}
\def\L{{\cal L}}
\def\s{s_\theta}
\def\st{s_{\tilde\theta}}
\def\c{c_\theta}
\def\ct{c_{\tilde\theta}}
\def\gt{\tilde g}
\def\et{\tilde e}
\def\At{\tilde A}
\def\Zt{\tilde Z}
\def\Wpt{\tilde W^+}
\def\Wmt{\tilde W^-}

\section{Introduction }
\label{section1}
Higgsless models
\cite{Csaki:2003dt,Agashe:2003zs,Csaki:2003zu,Nomura:2003du}
\cite{Barbieri:2003pr,Burdman:2003ya,Cacciapaglia:2004jz,Davoudiasl:2004pw,Barbieri:2004qk},
 which  have been proposed as
gauge theories in extra dimensions as an alternative to the standard
electroweak symmetry breaking mechanism, can also be understood as
four dimensional deconstructed theories in the context of linear
moose models
\cite{Arkani-Hamed:2001ca,Arkani-Hamed:2001nc,Hill:2000mu,Cheng:2001vd}
\cite{Son:2003et,
Foadi:2003xa,Hirn:2004ze,Casalbuoni:2004id,Chivukula:2004pk,Chivukula:2004af,
Georgi:2004iy,Perelstein:2004sc,Casalbuoni:2005rs,Foadi:2005hz}.
Deconstruction is a useful tool, both for computing and also for
finding renormalizable extensions \cite{Georgi:2005dm}.

One of the interesting features of the Higgsless
models is the possibility to delay  the unitarity violation scale
via the exchange of massive KK modes
\cite{Csaki:2003dt,SekharChivukula:2001hz,
Chivukula:2002ej,Chivukula:2003kq,DeCurtis:2002nd,DeCurtis:2003zt,
Abe:2003vg,Papucci:2004ip}. However, in   the simplest versions of
these models,  it is difficult to reconcile a delayed unitarity with
the electroweak constraints: in fact the $\epsilon_3$ parameter
 tends to get a large contribution. A recent solution to the $\epsilon_3$
problem, which does not spoil the unitarity requirement at low
scales, has been found by delocalizing the fermions along the fifth dimension
\cite{Cacciapaglia:2004rb,Foadi:2004ps,Foadi:2005hz}.
This solution has also a deconstructed correspondence, which
is obtained by introducing direct couplings between
ordinary left-handed fermions and the gauge vector bosons along the
moose string \cite{Casalbuoni:2005rs}. The direct coupling is realized in terms of a product
of non linear $\sigma$-model scalar fields that in the continuum
limit becomes a Wilson line. Also the fermion mass term in the deconstructed model is obtained by a
Wilson line connecting the left end of the chain with the right end.
The solution to
the $\epsilon_3$ problem is found by fine tuning the contribution from the gauge fields with the one from the fermions.
This cancellation may be ''local''
that means site by site \cite{Casalbuoni:2005rs,SekharChivukula:2005xm} or global.
The possibility of this cancellation was already noticed in
\cite{Anichini:1994xx} within a strongly interacting electroweak framework
 which can be reduced to a linear moose model.
In the continuum limit this solution
is obtained by allowing the left-handed fermion
fields to have some finite extent in the extra dimensions \cite{Cacciapaglia:2004rb,Foadi:2004ps}.
A different solution, suggested by holografic QCD, has been recently proposed
\cite{Hirn:2006nt}, where different metrics are felt by axial and vector
states.

Aim of this paper is to
show how the
direct couplings of the fermions to the bulk gauge bosons
defined in \cite{Casalbuoni:2005rs} can be
effectively obtained by integrating over the bulk fermions.
This result is obtained in the deconstructed theory. By performing the continuum limit we get an extension of
 the analysis of
\cite{Foadi:2004ps,Foadi:2005hz} to a generic warped case.

Our paper is organized as follows: starting  with a 5 dimensional gauge theory in a generic warped metric we write
down the equivalent deconstructed moose. Then we study  the new physics effects
in the low energy limit by eliminating the fermions living on the internal sites with the solutions
of their equations of motion in the limit in which the kinetic terms are negligible.
In this way effective couplings of the standard fermions to the gauge fields along the moose are generated.
They give a contribution to $\epsilon_3$ which can cancel "locally" or "globally"
the gauge sector contribution. In addition, with the same mechanism, fermion masses are generated.
Going back to  the continuum limit,
we have shown that the implementation of local cancellation in any generic  warped metric
leaves massless fermions. This is due to the presence of one horizon on
the infrared (IR) brane. However we
 can  require a global cancellation of the new physics contributions to the $\eps_3$
parameter. This fixes relations among the warp factor and the parameters of the fermion
and gauge sectors.
As in the flat case \cite{Foadi:2004ps,Foadi:2005hz},  in the range of parameters
allowed by the electroweak constraints
and compatible with the unitarity requirement,
it is possible to achieve fermion mass values
up to the charm or to the bottom quark.
In order to give mass to the top quark, a possible solution, following  \cite{Cacciapaglia:2004rb,Foadi:2005hz},
is to break five dimensional Lorentz invariance in the fermion sector. This is done by  considering
the fermionic action
over a fifth dimension   of inverse length  $(\pi R)^{-1}$ and curvature $k$ rescaled by a  factor $\kappa$.
This procedure does not spoil the flavour universality avoiding problems with FCNC and preserves the
results for $\epsilon_3^N$ cancellation (with $\epsilon_3^N$ we indicate the new physics contribution to
$\eps_3$).
We show that the overall results are quite independent on the warping of the space.

In section \ref{section2} we consider  a continuum model describing a $SU(2)$ gauge theory in
five dimensions with $SU(2)$ symmetry broken to $U(1)$ by boundary
conditions on the  branes. Following \cite{Foadi:2005hz} we add kinetic terms on the branes both for
bosons and fermions in a suitable way to recover, at low energy, the
standard model (SM). After discretization we write down the equivalent moose.
In section \ref{section3} we consider the fermion action and we deconstruct it.
In section \ref{section4}
 we
study  the new physics effects in the low energy limit by decoupling the heavy fermions. This
is done by eliminating the non-standard model fields with their
equations of  motion in the limit in which the kinetic terms are
negligible.
In section \ref{section5} we study the possible cancellations in the $\eps_3^N$
parameter going back to the continuum.
Conclusions can be found in  section \ref{section6}.

\section{Review of the model: gauge sector}
\label{section2}

Let us consider a theory with one extra dimension
with arbitrary curvature which preserves
the Poincar\`e invariance in the usual four dimensions: \be
ds^{2}=e^{-2\phi(y)}dx^{\mu}dx^{\nu}\eta_{\mu\nu}-dy^{2} \label{eq1}
\ee
The function $e^{-2\phi(y)}$ is called the warp-factor. It is always possible to
perform a coordinate transformation $z=z(y)$ to get a conformally flat
metric. This is achieved by requiring the following relation between $dy$ and $dz$
\be
dy=e^{-A(z)} dz
\ee
such that
\be ds^{2}=e^{-2A(z)}(\eta_{\mu\nu}dx^{\mu}dx^{\nu}-dz^{2})
\label{eq2} \ee
with $A(z)=\phi(y)$.

Let us consider a $SU(2)$ gauge theory in the 5-dimensional bulk,
with additional brane kinetic terms  to lower the mass values of the lightest gauge bosons,
and with boundary conditions which
break the gauge group to the electromagnetic $U(1)$.
Let us generalize the action in \cite{Foadi:2004ps} on a curved background, with a compact fifth dimension
varying on a segment of length $\pi R$.
The gauge action is: \bea
&&S_{gauge}=-\frac{1}{4}\int
d^4 x dz \sqrt{|g|} \frac{1}{g^2_{5}(z)} [G_{MN}^aG_{PQ}^a g^{MP}g^{NQ}]-\nn\\
 &&-\frac{1}{4}\int d^4 x dz \sqrt{|g|} \frac{\delta (z)}{\gt^{2}}
[G_{\mu\nu}^a G_{\sigma\rho}^a g^{\mu\sigma}g^{\nu\rho}] -\frac{1}{4}\int
d^4 xdz \sqrt{|g|} \frac{\delta (\pi R-z)}{{\gptt}}[
G_{\mu\nu}^3 G_{\sigma\rho}^3 g^{\mu\sigma}g^{\nu\rho}]\label{eq3}
\eea
with
\be
G_{MN}=\partial_M V_N-\partial_N V_M-i [V_M,V_N]\label{eq4}
\ee
where $V_M=V_M^a\tau^a/2$, $M=(5,\mu)$, and $a=1,2,3$ is the $SU(2)$ index. We have included a $z$-dependence in
the gauge coupling constant. We will comment on that later on.
We impose  the Dirichlet boundary conditions $V^{1,2}_\mu\vert_{z=\pi R}=0$, and the
 Neumann boundary conditions $\partial_z V^a_\mu\vert_{z=0}=0$ \cite{Foadi:2004ps}.

By substituting $\sqrt{|g(z)|}=e^{-5A(z)}$  we get:
\bea
S_{gauge}=&&-\frac{1}{4}\int d^4 xdz\frac{1}{g^2_{5}(z)} e^{-A(z)}[G^a_{\mu\nu}G^{a\mu\nu}-2G^a_{\mu
5}G^{a \mu 5}]-\frac{1}{4}\int d^4 xdz \frac{\delta (z)}{\gt^{2}}
e^{-A(z)} G^a_{\mu\nu}G^{a\mu\nu}\nn\\&&-\frac{1}{4}\int d^4 xdz
\frac{\delta (\pi R-z)}{\gptt} e^{-A(z)}G^3_{\mu\nu}G^{3\mu\nu}\label{eq5}
\eea

\begin{figure}[h] \centerline{
\epsfxsize=12cm\epsfbox{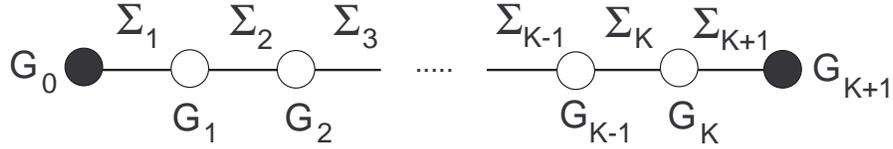} } \caption {{\it  The linear
moose model. \label{fig:1} }}
\end{figure}

Let us now review the deconstruction procedure of a gauge theory in 5 dimensions
\cite{Arkani-Hamed:2001ca,Arkani-Hamed:2001nc,
Hill:2000mu,Cheng:2001vd,Randall:2002qr}. In order to discretize the fifth dimension and to write the moose action which,
in the continuum limit corresponds to eq.~(\ref{eq5}),
 let us divide the $z$-segment in $K+1$ intervals of size $a$. In general we can consider different $a_j=z_{j+1}-z_j$
 lattice spacings near site $j$, but, since we are working with a general warped metric with the only
 requirement of Poincar\`e invariance
 (flat branes), we can safely consider equal spacing $a_j=a$ without loosing generality.
 The continuum limit is obtained
by taking $a\to 0$ and $K\to\infty$. Through discretization of the fifth dimension we get a finite
set of 4-dimensional gauge theories,
each of them acting at a particular lattice site.  In this way the discretized version of the original
5-dimensional gauge theory is
substituted by a collection of four-dimensional gauge theories synthetically described by a moose diagram
(see Fig. \ref{fig:1}).

\bea
S_{gauge}^{lattice}&&=-\frac{1}{2}\int d^4 x[\sum _{j=1}^{K}
\frac{a}{g_{5j}^{2}}{\rm Tr}(G_{\mu\nu}^{j})^{2}e^{-A_{j}}-2\sum
_{j=1}^{K+1}\frac{a}{g_{5j}^{2}}e^{-A_{j}}{\rm Tr}(G_{\mu 5}^{j})^{2}]\nn\\
&&-\frac{1}{2}\int d^4 x \frac{1}{\gt^{2}}e^{-A_{0}}{\rm
Tr}(G_{\mu\nu}^{0})^{2}-\frac{1}{2}\int d^4 x
\frac{1}{\gptt}e^{-A_{K+1}}{\rm
Tr}(G_{\mu\nu}^{K+1})^{2}\label{moose}\eea where, for sake of
simplicity, we have omitted the  gauge  index; $G_{\mu\nu}^{j}$ are
the 4-dimensional field strengths on the sites and
\be G_{\mu
5}^{j}=\partial_{\mu}V_{5}^{j}-\frac{\displaystyle 1}{\displaystyle
a}(V_{\mu}^{j}-V_\mu ^{j-1})-i[V_\mu^j,V_5^j] ~~~~~~~ j=1,..,K+1\ee

Here $V_\mu^j=V_\mu^{ja}\tau^a/2$ and $g_{5j}$ are the gauge fields and gauge coupling
constants associated to the groups $G_j$, $j=1,\cdots ,K$,
and $V_\mu^0=\Wt_\mu^{a}\tau^a/2$, $V_\mu^{K+1}=\yyt\tau^3/2$, are the gauge fields associated
to $SU(2)_L$ and $U(1)_Y$ respectively.

Let us introduce the link variables and their covariant derivatives for the discretized extra dimension:
\bea \Sigma_j=&&e^{-iaV_{5}^{j}}~~~~j=1,..,K+1\nn\\
D_\mu\Sigma_j=&&\partial_{\mu} \Sigma_j -iV_{\mu}^{j-1}\Sigma_j+i \Sigma_j V_{\mu}^{j}\label{unit2}\eea

For small lattice spacing,
\be
D_\mu\Sigma_j\sim -i (a\partial_\mu V_5^j-V_{\mu}^{j-1}+V_{\mu}^{j}
-ia[V_\mu^{j-1},V_5^j])\sim 
-ia G_{\mu 5}^j
\label{unit3}\ee
and so:
\be
(G_{\mu 5}^{j})^{2}\sim \frac{(D_{\mu}\Sigma_{j})^{\dag}(D^{\mu
}\Sigma_{j})}{a^{2}}\label{unit4}
\ee
from which the action for the discretized theory, in the small spacing limit, follows

\bea
S_{gauge}^{lattice}=&& -\frac{1}{2}\int d^4 x [\sum
_{j=1}^{K}\frac{a}{g_{5j}^{2}}e^{-A_{j}}{\rm Tr}(G_{\mu\nu}^{j})^{2}-2\sum
_{j=1}^{K+1}\frac{a}{g_{5j}^{2}}\frac{e^{-A_{j}}}{a^{2}}{\rm Tr}(D_{\mu}\Sigma_{j})^{\dag}(D^{\mu
}\Sigma_{j})]\nn\\
&&-\frac{1}{2}\int d^4 x
\frac{1}{\gt^{2}}e^{-A_{0}}{\rm Tr}(G_{\mu\nu}^{0})^{2}-\frac{1}{2}\int d^4 x
\frac{1}{\gptt}e^{-A_{K+1}}{\rm Tr}(G_{\mu\nu}^{K+1})^{2}\label{unit5}\eea

This action can be related to a linear moose action, based on the $SU(2)$ symmetry and
written in terms of $K+1$ non linear $\sigma$-model scalar fields $\Sigma_i$, ${i=1,\cdots ,K+1}$,
$K$ gauge groups, $G_i$, ${i=1,\cdots ,K}$,
 a global symmetry $SU(2)_L\otimes SU(2)_R$, in which   the standard electroweak
gauge group
  $SU(2)_L\times U(1)_Y$ is obtained by gauging
a subgroup of $SU(2)_L\otimes SU(2)_R$. The moose action  has the following form:
\be
S_{gauge}^{moose}=\int d^4 x[\sum
_{j=1}^{K+1}\frac{f_{j}^{2}}{g_{j}^{2}}{\rm Tr}[D_{\mu}\Sigma_{j}^{\dag}D^{\mu
}\Sigma_{j}]-\frac{1}{2}\sum
_{j=0}^{K+1}\frac{1}{g_{j}^{2}}{\rm Tr}(G_{\mu\nu}^{j})^{2}]\label{mo}\ee

 Therefore, by comparing eq.~(\ref{unit5}) with eq.~(\ref{mo}), we have a matching between the 5D
 parameters of the discretized theory
 (the gauge coupling constants $g_{5j}$, the lattice spacing $a$, the warp factors $A_j$, the gauge
 couplings $\gt$ and $\gpt$ on the branes)
 and the parameters of the 4D
 deconstructed theory (the gauge couplings along the chain $g_j$, the link couplings $f_j$, the gauge
 couplings at the left and right ends of the chain
 $g_0$ and $g_{K+1}$). Namely
 \bea
 \frac{ae^{-A_{j}}}{g_{5j}^{2}}&\longleftrightarrow& \frac{1}{g_{j}^{2}}~~~~~~~j=1,..,K\nn\\
\frac{e^{-A_{j}}}{ag_{5j}^{2}}&\longleftrightarrow& \frac{f_{j}^{2}}{g_{j}^{2}}~~~~~~~j=1,..,K+1\nn\\
\frac {e^{-A_{0}}} {\gt^2} &\longleftrightarrow&\frac{1}{g^2_{0}}\nn\\
\frac {e^{-A_{K+1}}}{\gptt}&\longleftrightarrow& \frac{1}{g^2_{K+1}}
\label{matching}\eea

In order to compare with the moose lagrangian given in ref. \cite{Casalbuoni:2005rs}
we only have to replace $V_\mu^j\rightarrow g_j V_\mu^j$ and $f_j\rightarrow f_j g_j$. Let us perform these rescalings for all $j=0,..,K+1$.
Notice that, since the mass dimension of the gauge couplings is 
 $[g_{5j}^{2}]=-1$, then $[V_{j}]=1$, that is $V_j$ have the canonical 4D mass dimension, and $g_j$ are dimensionless.

From the first two relations in eq.~(\ref{matching}), we obtain that, for all the sites
\be
a^{2}=\frac{1}{f_{j}^{2}g_{j}^{2}}
\label{const}
\ee
this means that, for generic $f_j$, it must be $g_j=\frac{1}{a f_j}$ and $g_{5j}^2=\frac{e^{-A_j}}{a f_j^2}$.

In the following, we will consider  two possibilities:
\begin{itemize}

\item{ $f_{j}=\bar f$ and $g_{j}=\bar g$ do not depend on $j$. In that case
the gauge coupling constants  depend on the site through: $g_{5j}^{2}=\bar g^{2}ae^{-A_{j}}$. The flat case corresponds to $e^{-A_{j}}=1$;}

\item{$g_{5j}=g_5$ does not depend on $j$, then
$f_{j}$ e $g_{j}$ are not constant; in particular
$g_{j}^{2}=\frac{g_{5}^{2}e^{A_{j}}}{a}$ and
$f_{j}^{2}=\frac{e^{-A_{j}}}{ag_{5}^{2}}$}.
\end{itemize}

\section{Review of the model: fermion sector}
\label{section3}

Proceeding as for  the gauge sector, let us consider fermions propagating in the
warped 5-dimensional space with additional  kinetic terms localized on the boundaries of the
fifth extra dimension:

\bea S_{ferm.}=\int d^4 xdz \sqrt{|g|}&&[ (\frac{i}{2}
\bar{\psi}e_{a}^{M}\Gamma ^{a}{\tilde D}_{M}\psi +
h.c.)-M\bar{\psi}\psi \nn\\
&& +\frac{\delta (z)}{{t}_{L}^{2}}i\bar{\psi}e_{a}^{\mu}\Gamma
^{a}{\tilde D}_{\mu}\psi+\delta (\pi
R-z)i\bar{\psi}(\frac{1}{t_R^2})e_{a}^{\mu}\Gamma ^{a}{\tilde
D}_{\mu}\psi ]
\label{sferm} \eea
where $e^{M}_{a}$ are the inverse
f\"unfbein defined by $g^{MN}=e_{a}^{M}e_{b}^{N}\eta ^{ab}$. In the
conformally flat metric defined by eq.~(\ref{eq2}) we have
$e^{M}_{a}=e^{A(z)}\delta ^{M}_{a}$. The five dimensional Dirac
matrices are defined in terms of the 4D ones by $\Gamma ^{a}=(\gamma
^{\mu},-i\gamma^{5})$. This is a generalization in warped space of
the fermion sector in \cite{Foadi:2004ps}.

The 5D fermions are equivalent to 4D Dirac fermions, $\psi=(\psi_L,\psi_R)$, where $\psi_L$ and
$\psi_R$ are $SU(2)$ left and
right-handed doublets for each family.
The boundary conditions we impose for the bulk fermions are: $ \psi_R\vert_0 =0,~\psi_L\vert_{\pi R}=0 $.
A sum over the flavors is implicit and the couplings ${t}_L$, ${t}_R$ can in general be different for each flavor.
Following \cite{Foadi:2004ps} we assume  an universal $t_L$, while $t_R$ will be fixed for each flavor in order
to reproduce the fermion mass spectrum. In eq.~(\ref{sferm}), $(1/t^2_R)$ is to be understood as a
$2\times 2$ diagonal matrix
with different entries for up and down fermions for each family:
\be
(\frac{1}{t_R^2})=\left(
\begin{array}{cc}
{({1/t^u_{R}})^{2}} & 0\\
0 & {({1/t^d_{R}})^{2}}\\
\end{array}\right)
\label{matrix}
\ee

The parameters ${t}_{L,R}$ set the weight of the brane kinetic terms with respect to the bulk one.
They parameterize the amount of extra dimension
which is felt by the fermions on the two branes. That is the delocalization in the bulk of the brane fermions.
If we do not include brane kinetic terms, the boundary conditions considered, imply the absence of a zero
mode for fermions
\cite{Csaki:2005vy,Foadi:2004ps}, and
the mass of the KK excitations is set to the only mass scale in the bulk, $1/R$. On the contrary,
in the limit ${t}_{L,R}\to 0$
the connection through the bulk kinetic terms is negligible and the model describes massless left-handed
fermions gauged under a $SU(2)\otimes U(1)$ gauge group living on the left brane, and massless
right-handed fermions gauged
under a $U(1)$ living on the right brane. As we will see, the masses of the SM fermions depend on
these two parameters. In fact the bulk fermions
make the communication between the light states possible, generating their masses.

In eq.~(\ref{sferm}),
 ${\tilde D}_{M}$  is the covariant derivative
${\tilde D}_{M}=D_{M}+\frac{1}{2}w_{M}^{bc}\sigma _{bc}$ where
$w_{M}^{bc}$ is the spin connection
\be
w_{M}^{ab}=\frac{1}{2}g^{RP}e_{R}^{[a}\partial
_{[M}e^{b]}_{P]}+\frac{1}{4}g^{RP}g^{TS}e_{R}^{[a}e_{T}^{b]}\partial
_{[S}e_{P]}^{c}e_{M}^{d}\eta _{cd}\ee
with  $\sigma _{ab}=\frac{1}{4}\gamma
_{[a}\gamma _{b]}$,  and
\be
D_{M}\psi=(\partial _{M}-i T^a V_{M}^{a}(z)-iY_{L}V^3_{M}(\pi
R))\psi\label {cova1}
\ee
where $T^a=\tau^a/2$ and $Y_L$ is the left hypercharge. Due to the additional kinetic terms, the 4D part of the
covariant derivatives on the left and right branes are:
\bea
D_{\mu}\psi_L\vert_{z=0}&=&(\partial _{\mu}-i T^a V_{\mu}^{a}(0)-iY_{L}V^3_{\mu}(\pi
R))\psi_L\nn\\
D_{\mu}\psi_R\vert_{z=\pi R}&=&(\partial _{\mu}-iY_{R}V^3_{\mu}(\pi R))\psi_R\label{cova}
\eea
where $Y_R=T^3+Y_L$ is the right hypercharge.

It is straightforward to show that, when the background geometry is
conformally flat with a conformal factor which depends only on the
fifth coordinate $z$, as given by eq.~(\ref{eq2}), the spin connection
contributions cancel each other. We get
\bea
&&S_{ferm.}=\int d^4 xdz
e^{-4A(z)}[(\frac{i}{2}\bar{\psi}\Gamma
^{M} D_{M}\psi + h.c.)-M(z)\bar{\psi}\psi]\nn\\
&&+\int
d^4 xdz e^{-4A(0)}\frac{\delta(z)}{t^{2}_{L}}i\bar{\psi}_L \gamma^{\mu}{D_{\mu}}\psi_L+
\int d^4 xdz e^{-4A(\pi R)}\delta(\pi
R-z)i\bar{\psi}_R (\frac{1}{t^{2}_{R}})\gamma^{\mu}{D_{\mu}}\psi_R\label{fer}
\eea
where we have defined
\be
M(z)=Me^{-A(z)}
\label{eq:21}
\ee

As for the gauge sector, let us discretize the fifth dimension. 

We will include in the action, following
\cite{Foadi:2005hz},
a $\kappa$ parameter in front of the $D_5$ term to be able to get the top quark mass. As we will see,
the fermion masses arise from the $\partial_5$ terms which mix left-handed and right-handed fermions.
The factor $\kappa$ will enhance the value of the fermion masses.
We get (in the unitary gauge, $V_5=0$):

\bea
S_{ferm.}^{lattice}&=&\int d^4 x[i\sum
_{j=1}^{K}ae^{-4A_{j}}\bar{\psi}_{j}\gamma^{\mu}{D_{\mu}}\psi_{j}+\frac{\kappa}{2}\sum
_{j=0}^{K}ae^{-4A_{j}}\bar{\psi}_{j}\gamma^{5}(\frac{\psi_{j+1}-\psi_{j}}{a})\nn\\
&-&\frac{\kappa}{2}\sum
_{j=0}^{K}ae^{-4A_{j}}(\frac{\bar{\psi}_{j+1}-\bar{\psi}_{j}}{a})
\gamma^{5}\psi_{j}-\sum
_{j=1}^{K}ae^{-4A_{j}}M_{j}\bar{\psi}_{j}\psi_{j}\nn\\
&+&\frac{i}{t^{2}_{L}}e^{-4A(0)}\bar{L}_{0}\gamma
^{\mu}{D_{\mu}}L_{0}+i e^{-4A(\pi R)}\bar{R}_{K+1}(\frac{1}{t^{2}_{R}})\gamma^{\mu}{D_{\mu}}R_{K+1}]+S^{Wilson}
\label{latt}\eea
where $\psi_j$, $j=0,..,K+1$ are the fermions at the $j$-site,
and  we have introduced the notation $(\psi_0)_L=L_0$, $(\psi_{K+1})_R=R_{K+1}$;
as before $a$ is the lattice spacing.
In eq.~(\ref{latt}), the 4D covariant derivatives acting on the $j$-site fermion has the form:
\bea
&&{D_{\mu}}\psi_{j}=(\partial
_{\mu}-i g_j T^{a}V_{\mu}^{aj}-i\gpt Y_{L}V_{\mu}^{K+1})\psi_{j}\nn\\
&&{D_{\mu}}\psi_{0}=(\partial
_{\mu}-i \gt V_{\mu}^{0}-i \gpt Y_{L}V_{\mu}^{K+1})\psi_{0}\nn\\
&&{D_{\mu}}\psi_{K+1}=(\partial
_{\mu}-i\gpt Y_{R}V_{\mu}^{K+1})\psi_{K+1}\label{covalat}
\eea
where, consistently with the moose gauge sector, we have rescaled $V_{\mu}^j\rightarrow g_j V_{\mu}^j$ with $g_0=\gt$
and $g_{K+1}=\gpt$.
To the lattice fermion action, we have added  a  Wilson term \cite{Cheng:2001vd}:
\be
S^{Wilson}=a\frac{\kappa}{4}\sum _{j=0}^{K}a e^{-4A_{j}}[\bar{\psi}_{j}
(\frac{\psi_{j+1}+\psi_{j-1}-2\psi_{j}}{a^{2}})+h.c.]\label{wilson}
\ee

Summing up:
\bea
S_{ferm}^{lattice}&=&\int d^4 x[i\sum _{j=1}^{K}ae^{-4A_{j}}\bar{\psi}_{j}\gamma^{\mu}
{D_{\mu}}\psi_{j}+\kappa \sum
_{j=0}^{K+1}\frac{e^{-4A_{j}}}{2}(\bar{\psi}_{j}(\frac{1+\gamma^{5}}{2})\psi_{j+1}\nn\\
&+&\bar{\psi}_{j}(\frac{1-\gamma^{5}}{2})
\psi_{j-1}
+\bar{\psi}_{j+1}(\frac{1-\gamma^{5}}{2})\psi_{j}+\bar{\psi}_{j-1}(\frac{1+\gamma^{5}}{2})\psi_{j})\nn\\
&-&\sum
_{j=1}^{K}e^{-4A_{j}}(aM_{j}+\kappa)\bar{\psi}_{j}\psi_{j}+\frac{i}{t^{2}_{L}}e^{-4A_0}\bar{L}_{0}\gamma^{\mu}
{D_{\mu}}L_{0}\nn\\
&+&i e^{-4A_{K+1}}\bar{R}_{K+1} (\frac{1}{t^{2}_{R}}) \gamma^{\mu}
{D_{\mu}}R_{K+1}]
\label{wilson1}\eea
Introducing the shorthand notation:
$L_{j}=\psi_{L}^{j}$ e $R_{j}=\psi_{R}^{j}$, and using the property that, in the continuum limit for the case
of a continuous metric, $e^{-4A_{j}}\simeq
e^{-4A_{j+1}}$, we get
\bea
S_{ferm}^{lattice}&=&\int d^4 x[i\sum _{j=1}^{K}ae^{-4A_{j}}\bar{\psi}_{j}\gamma^{\mu}
{D_{\mu}}\psi_{j}+\kappa\sum
_{j=0}^{K}e^{-4A_{j}}(\bar{L}_{j}R_{j+1}+h.c.)\nn\\
&-&\sum
_{j=1}^{K}e^{-4A_{j}}(aM_{j}+\kappa)(\bar{L}_{j}R_{j}+h.c.)+\frac{i}{t^{2}_{L}}e^{-4A_0}\bar{L}_{0}\gamma^{\mu}{D}_{\mu}
L_{0}\nn\\&&+
i e^{-4A_{K+1}}\bar{R}_{K+1}(\frac{1}{t^{2}_{R}})\gamma^{\mu}{D}_{\mu}R_{K+1}]\label{wilson2}
\eea

It is convenient to rescale the fermion fields in order to make them canonical in 4 dimensions.
We define
\bea
\hat{\psi}_j&=&  \sqrt{a}  e^{-2 A_j}\psi_j,~~~~~~~~ j=1,..,K\nn\\
\hat{L}_{0}&=& \frac{e^{-2 A_0}}{t_L}{L}_{0}\nn\\
\hat{R}^{u,d}_{K+1}&=& \frac{e^{-2 A_{K+1}}}{t^{u,d}_R}{R}^{u,d}_{K+1}\label{ridef}
\eea
where we have used the notation $\psi=\left(\begin{array}{c} \psi^u\\\psi^d\end{array}\right)$ for each family and
$t_R^{u,d}$ are the eigenvalues of the $t_R$ matrix (see eq.~(\ref{matrix})).

 With the redefinitions of eq.~(\ref{ridef}) we get:
\bea
&&S_{ferm.}^{lattice}=\int d^4 x[\sum
_{j=1}^{K}i\bar{\hat{\psi}}_{j}\gamma^{\mu}{D}_{\mu}\hat{\psi}_{j}
+i\bar{\hat{L}}_{0}\gamma^{\mu}
{D}_{\mu}\hat{L}_{0}+i\bar{\hat{R}}_{K+1}\gamma^{\mu}
{D}_{\mu}\hat{R}_{K+1}\nn\\
&&-\frac{1}{a}\sum
_{j=1}^{K}(aM_{j}+\kappa)(\bar{\hat{L}}_{j}\hat{R}_{j}+h.c.)
+\frac{\kappa}{a}
\sum
_{j=1}^{K-1}(\bar{\hat{L}}_{j}\hat{R}_{j+1}+h.c.)\nn\\
&&+ \frac{t_L}{\sqrt{a}}\kappa(\bar{\hat{L}}_{0}\hat{R}_{1}+h.c.)+
\sum_{f=u,d}\frac{t^f_R}{\sqrt{a}}\kappa(\bar{\hat{L}}^f_{K}\hat{R}^f_{K+1}+h.c.)]
\label{ridef1}
\eea
This action describes fermions on the $j$-sites with $j=0,..,K+1$, with a mass term
\be
m_j=(a M_j+\kappa)/a,~~~~ j=1,..,K
\label{mj}\ee
which "hop" from one site to the near one. These "hopping" terms come from the derivative
term along the fifth dimension after discretization.
The "hopping" strengths are:
\be
\alpha_0=\kappa~ t_L/\sqrt{a}
\label{a0}
\ee
which parameterizes the probability for the fermions on
the left end of the moose to hop to the $j=1$ site;
\be
\alpha^{u,d}_K =\kappa~ t^{u,d}_R/\sqrt{a}
\label{aK}
\ee
parameterizing the probability for the fermions on
the right end of the moose to hop to the $j=K$ site, and
\be
\alpha_j =\kappa/a
\label{aj}
\ee
 parameterizing the probability for the fermions
on the $(j-1)$-site to hop to the $j$-site. Notice that all the $\alpha_j$ are equal for $j=1,..,K-1$.

\section{Decoupling the heavy fermions}
\label{section4}

Let us study the effects of the $\psi_j$ ($i=1,\dots,K$) fermions in the low-energy limit, that is for  kinetic terms negligible with respect to mass terms.

This can be done by eliminating the $\psi_j$ fields with the solutions
of their equations of motion.
Actually we want to derive the trilinear effective interactions between the light fermions living
on the left and right ends of the moose
with the gauge bosons on the $j$-sites. These effective vertices are provided by the mixing between the
light fermions and
the heavy ones living on the $j$-sites. This means that the contributions to the effective interactions
could come only from the
quadratic interactions among fermions described in eq.~(\ref{ridef1}). For this reason, in solving the
equations of motions,
it is enough to consider the quadratic part of the fermionic action.

Let us solve the equations of motion for the fields ${\hat L}_j$, $(j=1,..,K)$ and
${\hat R}_j$, $(j=1,..,K)$ in terms of $\hat L_0$ and $\hat R_{K+1}$
which respectively are the left and right components of the SM fermions.
The equations of motion, from the quadratic part of the action, and neglecting the kinetic term contributions are:
\bea
&&\alpha_{j}\hat{L}_{j}-m_{j+1}\hat{L}_{j+1}=0,~~~~j=0,..,K-1\nn\\
&&\alpha_{j}\hat{R}_{j+1}-m_{j}\hat{R}_{j}=0,~~~~~j=1,..,K\label{eqmotion}
\eea
The solutions  are:
\bea
&&\hat{L}_{1}=\frac{\alpha_0}{m_1}\hat{L}_{0};\nn\\
&&\hat{L}_{j}=(\frac{\alpha_0}{m_j}\prod_{i=1}^{j-1}\frac{\alpha_{i}}{m_{i}})\hat{L}_{0};~~~~~~~~~~~~j=2,..,K\nn\\
&&\hat{R}^{u,d}_{j}=(\frac{\alpha_K^{u,d}}{m_K}
\prod^{K-1}_{i=j}\frac{\alpha_{i}}{m_{i}})\hat{R}^{u,d}_{K+1}
~~~~~j=1,..,K-1\nn\\
&&\hat{R}^{u,d}_{K}=\frac{\alpha_K^{u,d}}{m_K}
\hat{R}^{u,d}_{K+1}
\label{solut}
\eea

By substituting in eq.~(\ref{ridef1}), we get:
\bea
&&S_{ferm.}^{lattice}=\int d^4 x
[i\bar{\hat{L}}_{0}\gamma^{\mu}
{D}_{\mu}\hat{L}_{0}+
\sum
_{j=1}^{K} b_j^L ~~i\bar{\hat{L}}_{0}\gamma^{\mu}
(\partial
_{\mu}-ig_j T^{a}V_{\mu}^{aj}-i\gpt Y_{L}V_{\mu}^{K+1})\hat{L}_{0}\nn\\
&&+i\bar{\hat{R}}_{K+1}\gamma^{\mu}
{D}_{\mu}\hat{R}_{K+1}+ \sum
_{j=1}^{K} \sum_{f=u,d}   b_j^{f R} ~~i\bar{\hat{R}}^f_{K+1}\gamma^{\mu}
(\partial
_{\mu}-i g_j T^{3}V_{\mu}^{3j}-i \gpt Y_{L}V_{\mu}^{K+1})\hat{R}^f_{K+1}
\nn\\
&&+\sum_{j=1}^{K}\sqrt{b_j^{Ru}}\sqrt{ b_j^{Rd}}\frac{g_j}{\sqrt{2}}(
\bar{\hat{R}}^u_{K+1}\gamma^{\mu}
V_{\mu}^{+j}\hat{R}^d_{K+1}+h.c.)
- \sum_{f=u,d}{\tilde m}^f
(\bar{\hat{L}}^f_{0}\hat{R}^f_{K+1}+h.c.)]
\label{fermion}
\eea
where ${D}_{\mu}\hat{L}_{0}$ and ${D}_{\mu}\hat{R}_{K+1}$ are given in eq.~(\ref{covalat}),
$V_{\mu}^{+j}=(V_{\mu}^{1j}-iV_{\mu}^{2j})/\sqrt{2}$,
 and
\bea
&&b_j^L=(\frac{\alpha_0}{m_j}\prod_{i=1}^{j-1}\frac{\alpha_{i}}{m_{i}})^2\nn\\
&&b_j^{Ru}=(\frac{\alpha^u_K}{m_K}\prod^{K-1}_{i=j}\frac{\alpha_{i}}{m_{i}})^2,
~~~~~b_j^{Rd}=(\frac{\alpha^d_K}{m_K}\prod^{K-1}_{i=j}\frac{\alpha_{i}}{m_{i}})^2\nn\\
&&{\tilde m}^{u,d}=\alpha_0 \frac{\alpha_K^{u,d}}{m_K}\prod^{K-1}_{i=1}\frac{\alpha_{i}}{m_{i}}\label{bj}
\eea

Notice that the following relations for the up and down-type fermion masses hold for all $j=1,..,K$:
\be
{\tilde m}^{u}=\sqrt{b_j^L}\sqrt{ b_j^{Ru}} m_j,~~~~{\tilde m}^{d}=\sqrt{b_j^L}\sqrt{ b_j^{Rd}} m_j
\label{mfsq}
\ee

The canonical kinetic terms for the standard fermions are obtained by the following redefinitions:
\bea
\hat L_0&\to& \frac 1{\sqrt{1+\sum_{i=1}^K b_i^L}}\hat L_0\nn\\
\hat R^f_{K+1}&\to& \frac 1{\sqrt{1+\sum_{i=1}^K b_i^{Rf}}}\hat R^f_{K+1}~~~~~~~f=u,d
\label{canonical}
\eea
so that the mass generated for the SM fermions is, $f=u,d$:
\be
m^f=m_j \sqrt{\frac {b_j^L} {1+\sum_{i=1}^K b_i^L} }
 \sqrt{\frac {b_j^{Rf}} {1+\sum_{i=1}^K b_i^{Rf}} }\,\,\,\,~~~~\forall j=1,..,K
 \label{massa}
\ee
The mass difference between, for example, the top and the bottom quark, can be obtained by
choosing $b_j^{R\,t}>>b_j^{R\,b}$. We will discuss this issue in more detail in Section \ref{section5}.

The action in eq.~(\ref {fermion}) can be  directly compared with the
moose model one given in ref. \cite{Casalbuoni:2005rs}, where
only standard model fermions were considered,  coupled to the SM gauge fields  at the ends of the chain.
Direct couplings of the left-handed fermions to the
fields $V_\mu^j$  were introduced by generalizing the procedure suggested in
the BESS model
\cite{Casalbuoni:1985kq,Casalbuoni:1986vq}. For each $\psi_L$,  the following $SU(2)$ doublets were constructed
 \be \chi^j_L=\Sigma_j^\dagger
\Sigma_{j-1}^\dagger\cdots \Sigma_1^\dagger
\psi_L\,,\,\,\,\,\,\,\,j=1,\dots,K\,
\label{eq:8} \ee

Therefore a  term containing direct left-handed fermion couplings to $V^j_\mu$,
invariant under the symmetry transformation of the model, could be added.
In the unitary gauge $\Sigma_j=I$,  therefore additional couplings between left-handed fermions and the gauge bosons
on the $j$-sites
were generated. They are exactly the ones in eq.~(\ref{fermion}) with strength $b^L_j$. In ref. \cite{Casalbuoni:2005rs}
these additional couplings
were described by free parameters; in the present derivation, $b^L_j$ are generated by the presence of heavy fermions
in the bulk. Notice that the sign
of $b_i^L$ is  positive definite and, as we will show, this is the right sign to compensate for
the contribution of the gauge bosons to the parameter $\eps_3^N$.

With the same mechanism, additional couplings $b_j^R$ of the right-handed standard fermions to the $V_\mu^j$
gauge bosons are generated. According to our choice of different values of $t_R$ for up and down-type fermions,
the couplings of the right-handed fermions to the heavy gauge bosons along the moose are different for up
and down components and for the various fermion families.

The effects of the $V_{\mu}^j~(j=1,...,K)$
particles in the low-energy limit  can be considered  by eliminating the $V_{\mu}^j$ fields with the solution
of their equations of motion for large gauge
coupling constants $g_j$. This limit
corresponds to heavy masses for the $V^j$ fields. In fact in this limit the kinetic terms of the new resonances
are negligible.
The corresponding effective theory will be considered up to order $(1/g_j)^2$.

By separating
charged and neutral components, we get
\be V_j^\pm=\frac 1
{g_j}(\gt\Wt^\pm z_j) \label{VCC}\ee
\be V_j^3=\frac 1 {g_j}(\gpt \yyt y_j+\gt\Wt^3z_j)
\ee

Here we are neglecting the fermion current contributions which give  current-current
interactions. These terms turn out to be quadratic in the $b_j^{L,R}$ parameters.

Coming back to the standard notation: $\hat L_0= \frac{1-\gamma_5}{2} \hat \psi$,
and $\hat R_{K+1}=\frac{1+\gamma_5}{2} \hat \psi$, we get the following effective interaction lagrangians:

\bea
\L_{eff}^{charged} &=& \frac{\et}{\sqrt{2} \st}\Big\{
\big(1-\frac{{\textsf b^L}}{2}\big)\bar{\hat \psi}_d
     \gamma^\mu\frac{1-\gamma_5}{2}{\hat \psi}_u \Wmt_\mu\nn\\
     &&+\frac{1}{2}\sqrt{{\textsf b^{Ru}}{\textsf b^{Rd}}}~
     \bar{\hat \psi}_d
     \gamma^\mu\frac{1+\gamma_5}{2}{\hat \psi}_u \Wmt_\mu
     \Big\} +~h.c.\label{eq:21}
     \eea
     \bea
\L_{eff}^{neutral} &=& -\frac{\et}{\st \ct} \big(1-\frac{{\textsf b^L}}{2}\big)\bar{\hat \psi}
     \gamma^\mu\ T^3 \frac{1-\gamma_5}{2}{\hat \psi} \Zt_\mu\nn\\
     &&-\frac{\et}{\st \ct}\sum_{f=u,d}\frac{{\textsf b^{Rf}}}{2}
     \bar{\hat \psi}
     \gamma^\mu\ T^3 \frac{1+\gamma_5}{2}{\hat \psi} \Zt_\mu\nn\\
&&+\et\frac{\st}{ \ct} \bar{\hat \psi}
     \gamma^\mu Q
       {\hat \psi} \Zt_\mu - \et \bar{\hat \psi} \gamma^\mu Q {\hat \psi}
       \At_\mu\,,\label{eq:22}
\eea
with $\et = \tilde g \st=\tilde g'\ct$ and
\be
 {\textsf b^L}=2 \frac {\sum^K_{i=1} b_i^L y_i}{1+\sum^K_{i=1} b_i^L},~~~~~~~~~~~
{\textsf b^{Rf}}=2 \frac {\sum^K_{i=1} b_i^{Rf} z_i}{1+\sum^K_{i=1} b_i^{Rf}},~~~~f=u,d
\ee

Here fields and couplings are "tilded" because they need renormalization (see \cite{Casalbuoni:2005rs}).
From eqs. (\ref{eq:21}), (\ref{eq:22}) we see that the $b_j^{R}$ parameters give rise to charged and neutral
right-handed currents coupled to the SM gauge bosons. There are strong phenomenological constraints on the ${\textsf b^{Rf}}$ parameters,
coming for example, from
right-handed fermion  coupling to  charged $W$  contribution to the  $b\to s\gamma$ process
\cite{Larios:1999au} and to the $\mu$ decay \cite{Eidelman:2004wy}. Nevertheless, as it is clear from eq.~(\ref{bj}),
in order to generate a mass
term for the SM fermions, we need all the $\alpha_j\ne 0$ for $j=0,..,K$.

However we will make the assumption, $\alpha_j>>\alpha_0>>\alpha^f_K$ for $j=1,..,K-1$;
this means $b^L_j>>b^{Rf}_j$.
(An exception will be done for the top quark which will require $\alpha_K^{t}\sim \alpha_0$, that is
$t_R^{t}\sim t_L$
in order to obtain the physical value for $m_t$ (see Section \ref{section5}). This choice will not spoil the results obtained
since the top  does not enter in the new physics contribution to the $\eps_3$ parameter).

In this approximation we can proceed exactly as in ref. \cite{Casalbuoni:2005rs} concerning the low-energy limit
of the model, the fields and couplings
renormalization and the calculation of the electroweak $\epsilon_1^N$,  $\epsilon_2^N$ and $\epsilon_3^N$ parameters.

Following the same lines of ref. \cite{Casalbuoni:2005rs}, one can expand up to the first order in $b^L_j$
and neglect terms $O(b_j^L/g_j^2)$. Analogously
we neglect corrections coming from $b_j^{Rf}$. With this approximation, we can also neglect the contribution
from the effective four-fermion couplings. 

Finally the new physics contribution to the $\epsilon$ parameters is:
\bea
\eps_1^N &\simeq& 0\,,\nn\\
\eps_2^N &\simeq& 0\,,\nn\\
\eps_3^N &\simeq&\sum_{i=1}^{K}y_i(\f {e^2 }{\s^2 g_i^2} z_i-b^L_i)
\label{epsappr}\eea
where $e$ is the electric charge, $s_{\theta}$ is the sine of Weinberg angle defined by
$G_F=\sqrt{2} e^2/(8 s_\theta^2 c_\theta^2 M_Z^2)$, and
\be z_i=\sum^{K+1}_{j=i+1}x_j,~~~x_i=\frac{f^2}{f_i^2},~~~\frac
1{f^2}=\sum_{i=1}^{K+1}\frac 1{f_i^2},~~~\sum_{i=1}^{K+1}x_i=1,~~~
y_i=1-z_i\ee
with $f_i$ given in eq.~(\ref{matching}) and rescaled according to $f^2_i\rightarrow f^2_i g^2_i$.

The  expression for $\eps_3^N $ suggests that the additional fermion couplings  to $V_i$  proportional to
$b^L_i$ can compensate site by site for the
contribution of the tower of gauge vectors with the choice
\be
b_i^L=\frac {e^2}{\s^2 g_i^2} z_i,~~~~ \forall i=1,\cdots K
\label{mooselocal}
\ee
or the whole contribution from the gauge sector can cancel with the fermion one.

In order to explore in detail the various possible way to realize this compensation, let us perform the continuum limit.

\section{The continuum limit}
\label{section5}

In this section  we consider the continuum limit of the deconstructed model, obtained for $a\to 0$, $K\to\infty$
with the condition
$(K+1) a=\pi R$, which is length of the segment in the fifth dimension.
Let us define:
\be\lim_{a\to 0} \frac {b_j^{L,R}} a=b^{L,R}(z),~~~\lim_{a\to 0}
a{f_j^2}=f^2(z),~~~\lim_{a\to 0}a g_j^2=g_5^2(z)e^{A(z)}
\label{continuum}\ee
Using
the relations between the moose parameters and the ones of the 5D theory given in eq.~(\ref{matching}),  we obtain
\be
\frac{1}{f^2}=\int_0^{\pi R} \frac{dz}{f^2(z)}=\int_0^{\pi R} dz~
g_5^2(z) e^{A(z)}
\label{fz} \ee
From eq.~(\ref{bj}), using eqs.~(\ref{mj}),(\ref{a0}),(\ref{aj}),
we get  the continuum limit for
the additional  fermionic couplings, generated by the decoupling of the heavy fermionic modes in the bulk:
\be
b^L(z)=\lim_{a\to 0}\frac {\kappa^2 t_L^2}{(a M_j +\kappa)^2} {\rm exp} [\sum_{i=1}^{j-1}
\log \frac {\kappa^2}{(a M_j +\kappa^2)^2}]=
{{t}^{2}_{L}} e^{\displaystyle{{-\frac{2}{\kappa}\int^{z}_{0} dt M(t)}}}
\label{bz}
\ee
\be
b^{Rf}(z)=
({t}^{f}_{R})^2 e^{\displaystyle{{-\frac{2}{\kappa}\int^{\pi R}_{z} dt M(t)}}}~~~~~~f=u,d
\label{bz1}
\ee

Finally, using eqs. (\ref{massa}), (\ref{bj}),
(\ref{mj}), we get the following expression for the fermion masses (neglecting terms ${\cal O}(b^2)$):
\be
m^{f}=\kappa\sqrt{b^L(z)b^{Rf}(z)}= \kappa
\frac {T_{L}T_{R}^f}{\pi R}~e^{\displaystyle{{-\frac{1}{\kappa}\int^{\pi
R}_{0}M(z)}dz}} =\kappa\frac {T_R^f} {\sqrt{\pi R}}
\sqrt{b^L(\pi R)}\label{mf}\ee
where we have introduced the dimensionless parameters $T_L=t_L\sqrt{\pi R}$, $T_R^f=t_R^f\sqrt{\pi R}$
with $f=u,d$.

In order to have a suppressed $\eps_3^N$ parameter, we can require the contribution from the gauge sector
to cancel with the one from the fermion sector. This cancellation may be local (for each value of the
fifth coordinate) or
global, namely we can require that the two contributions, integrated over the $z$-coordinate, do cancel.

\subsection{Local cancellation in $\eps_3^N$}

 Let us start investigating the possibility of a local cancellation. We need the continuum limit of
 the following expression, appearing in eq.~(\ref{epsappr}),
\be
\lim_{a\to 0}\frac 1 a \f {e^2 }{\s^2 g_i^2} z_i=
\frac{e^2}{s_{\theta}^2} \frac{e^{-A(z)}}{g_5^2(z)}
\frac{\int^{\pi
R}_{z}dt { g_5^2(t) e^{A(t)}}}
{\int^{\pi R}_{0}dt g_5^2(t) e^{A(t)}}=\frac {e^2}{s_\theta^2}f^2 f^2(z)
\int_z^{\pi R}dt \frac {1}{f^2(t)}
\ee

Local cancellation requires:

\be
b^L(z)=\frac{{T}^{2}_{L}}{\pi R}~e^{\displaystyle{{-\frac{2}{\kappa} \int ^{z}_{0} dt
{M}(t)}}}= \frac {e^2}{s_\theta^2}f^2 f^2(z)
\int_z^{\pi R}dt \frac {1}{f^2(t)}
\label{local1}\ee

This equality must holds for each value of the fifth coordinate. The quantity on the right-hand site vanishes in
$z=\pi R$ (at least for continuous functions $f(z)$).
As a consequence $b_L(\pi R)=0$, and, according to eq.~(\ref{mf}), the fermion mass is zero:
if we impose the local cancellation of the $\epsilon_3^N$ parameter,
we can't give mass to
fermions.

As an example, let us consider  the first choice of parameters
 described    at the end of Section \ref{section2}.
This choice, ($f_{j}=\bar f$ and $g_{j}=\bar g$), leads to a $z$-dependent five dimensional gauge coupling constant,
with $g_5^2(z)e^{A(z)}$ independent on $z$ due to eq.~(\ref{continuum}).
Local cancellation requires:

\be
b^L(z)=\frac{{T}^{2}_{L}}{\pi R}~e^{\displaystyle{{-\frac{2}{\kappa} \int ^{z}_{0} dt
{M}(t)}}}=\frac{e^{2}}{s_{\theta}^{2}g_5^2(0)}(1-\frac{z}{\pi R})
\label{local}\ee
where we have normalized $e^{-A(0)}=1$.
In order to
satisfy eq.~(\ref{local}), we have to take: \bea
&& e^{A(z)}=1-\frac{z}{\pi R}\nn\\
&& M=\frac{\kappa}{2\pi R}~,~~~~~~~~~~~~t_L^2=\frac{T_L^2}{\pi R}=\frac{e^2}{\s^2 g_5^2(0)}\label{match}
\eea
With this choice for the metric, which turns out to be singular on the right brane, it is not possible
to give mass to fermions.
This follows from the fact that in our model the right handed fermions are on one horizon of the metric
and are causally disconnected from the left handed fermions.

This conclusion is general:  even if one can find a metric such to obtain a local cancellation of the
gauge and fermion contributions
to the $\epsilon_3^N$ parameter, this metric has a singularity on the right end brane which prevents the fermions
to acquire a mass.

\subsection{Global cancellation in $\eps_3^N$}

The other possibility is to require that the whole contribution to $\epsilon_3^N$ from
the gauge sector cancels with the one from fermions. That means \be
\epsilon_{3}^N=\int^{\pi R}_{0}dz \frac{z}{(\pi
R)^{2}}\Big[\frac{e^{2}}{s^{2}_{\theta}g^2_{4}(z)}(1-\frac{z}{\pi R})-
T^{2}_{L}e^{\displaystyle{{-\frac{2}{\kappa} \int^z_{0}dt {M}(t)}}}\Big]=0 \ee with
\be
g^2_4(z)=g_5^2(z)/(\pi R)\ee

For a flat metric, corresponding to the choice $e^{-A(z)}=1$, leading
to $M(z)=M={\hat M}/(\pi R)$ and $g_4(z)=g_4$, we get (this analysis
has already been performed by \cite{Foadi:2004ps})
\be
\epsilon_{3}^N=\frac{\lambda^2}{6}-T_{L}^{2} \hat A
\ee where
\be
\lambda =\frac{e}{s_{\theta}g_{4}}~~,~~~~~ \hat
A=\frac{\kappa^2}{4\hat{M}^{2}}-\frac{\kappa}{2\hat{M}}(1+\frac{\kappa}{2\hat{M}})e^{-2\hat{M}/\kappa}
\ee
The requirement $\epsilon_{3}^N=0$ links the parameter of the
gauge sector $\lambda$, with the fermion parameters $t_L$ and $M$
as found in \cite{Foadi:2004ps}.

We are now in the position to explore how these results can change by considering warped-metrics. Let us,
for example, specialize to a Randall-Sundrum metric:
\be
ds^{2}=\frac{1}{(1+kz)^{2}}(dx^{2}-dz^{2})\label{RS} \ee
 with $k$
the curvature along the $z$-coordinate.

The fermion mass, turns out to be:
\be
m^f= \kappa \frac{T_L T_R^f}{\pi R}(1+ k \pi R)^{-\frac{M}{k \kappa}}
\label{fmass}
\ee

Again, let us analyze  the two cases
 described    at the end of Section \ref{section2}.
\begin{itemize}
\item{
$f_{j}=\bar f$ and $g_{j}=\bar g$ do not depend on $j$;
$g^2_5(z)=g^2_5(0)/(1+kz)$.

\be
\epsilon_{3}^N=\frac{\lambda^2(0)}{6}-\frac{T^{2}_{L}}{k\pi
R}[\frac{(1+k\pi
R)^{-2\frac{M}{k \kappa}+1}}{(-2\frac{M}{k \kappa}+1)}
+\frac{1-(1+k\pi
R)^{-2\frac{M}{k \kappa}+2}}{k\pi R(-2 \frac{M}{k \kappa}+1)(-2\frac{M}{k \kappa}+2)}]\label{e3}
\ee
where we have used $g_{4}^{2}(z)(1+kz)=g_{4}^{2}(0)$ and introduced $\lambda(0)= e/({s_{\theta}g_{4}(0)})$.

Obviously  the contribution from the gauge sector is the same of the flat case,
in fact we have chosen the $z$ dependence of $g_5(z)$   in a way to compensate the warp factor.

Following the analysis in \cite{Casalbuoni:2005rs}, we can derive
the mass of the lightest charged gauge boson, the $W$, which, after fields and couplings renormalization,
in the limit of small $\lambda(0)$ is
 given by:
\be
M_{W}\sim\frac{\lambda(0)}{\pi
R}(1-\frac{\lambda^2(0)}{6})\ee

To have an estimate of the contribution coming from the fermion sector, we can consider the case
of large curvature: $k>>(\pi R)^{-1}, ~M$.
In this limit we get:
\be
\epsilon_{3}^N\sim\frac{\lambda^2(0)}{6} -\frac {T_{L}^{2}}{2}
\ee
which is the same result obtained for flat metric and $M\to 0$.
This was expected since, in this limit, all the mass scales are negligible with respect to the curvature $k$.
The warping factor is multiplied by $M$, so its effect is weakened for $M \to 0$.
 The global cancellation requires
\be
T_L\sim \f 1 {\sqrt{3}}\pi R M_W
\ee
This implies that  the fermion mass is:
\be
m^{f}\sim \kappa \frac{T_{R}^f}{\sqrt{3}} M_W\ee
 }

\item{ $g_5(z)=g_5$.

The expression for $\epsilon_3^N$ in this case is quite complicated. For
large curvature: $k>>(\pi R)^{-1}, ~M$ we get
\be
\epsilon_{3}^N\sim \frac  {1} {4 k\pi R} \lambda^2 -\frac{T_{L}^{2}}{3}
\label{eps3w}
\ee
and, to the first order in $\lambda$,
\be
M_{W}\sim\frac{\lambda}{\pi R}\sqrt{\frac{2}{k\pi R}}
\label{mww}
\ee

Both  the results given in eq.~(\ref{eps3w}) and eq.~(\ref{mww}) contain the factor
$\lambda_{eff}=\lambda/\sqrt{k\pi R}$. This is a
consequence of the following observation.  The bulk gauge
eigenstates are localized near the right brane of our model,
which corresponds to the IR brane in the Randall-Sundrum metric
given in eq.~(\ref{RS}). The bulk fields that connect the
$SU(2)_L$ and the $U(1)_Y$ theories  transmit the breaking down
to $U(1)_{em}$. Then  the phenomena responsible for the electroweak symmetry
breaking happen
mainly near the IR  brane. Since $g_5$ is a dimensional parameter,
it gets shifted by the warp factor $g_5^2\to g_5^2 /(1+k \pi R)$. In
the limit of large curvature this rescaling is responsible for the
extra factor in $\lambda^2_{eff}$.

In this case
the global cancellation requires
\be
T_L\sim \pi R \sqrt{\f 3 8} M_W
\ee
 a result which is very close to the flat case.
Analogously, for the fermion masses, we get
\be
m^{f}\sim \kappa \sqrt{\frac 3 8}{T_{R}^f}  M_W
\ee}
\end{itemize}

Therefore, in both  cases the numerical value of $T_L$, necessary to satisfy the electroweak constraints, is determined once $R$
is given. For instance choosing $R^{-1}\sim 1$ TeV, we have $T_L\sim 0.15$. A smaller $T_L$ requires a higher $R^{-1}$ which implies a partial wave unitarity
violation at a lower scale.

From eq.~(\ref{fmass}) 
we notice that the ratio $T^{t}_R/T^{b}_R=m_t/m_b$ 
does not depend on the parameters $M,k,\kappa$ and $R$. In addition the
product of $T^{t}_R T^{b}_R$ is constrained by the
limit on right-handed charged currents.

In this model a right-handed coupling $tbW$ is generated by eliminating the fermions in the bulk.
In eq.~(\ref{eq:21}), in the low energy limit, we can read the effective coupling of the standard $W$ boson
to the right-handed fermion current.
\be
{\cal L}^{tbW}= \frac{\tilde g}{\sqrt{2}}\kappa_R^{CC}\Big (
     \bar{\hat \psi}_b
     \gamma^\mu\frac{1+\gamma_5}{2}{\hat \psi}_t \Wmt_\mu
   + h.c.\Big)
\ee
with
\be
\kappa_R^{CC}=\frac{1}{2}\sqrt{{\textsf b^{R\,t}}{\textsf b^{R\,b}}}
    =\sum_{j=1}^K \sqrt{b_j^{R\,t}} \sqrt{b_j^{R\,b}} z_j
\ee

The tilded variables differ from the physical ones by corrections which are of the second order in $1/g_j^2$
and first order in $b_j^{L,R}$. Since the $tbW$ coupling is already of the first order in $b_j^{R}$, we can neglect
these additional  corrections.

By performing the continuum limit, we get
\be
\kappa_R^{CC}=\frac{T_R^{t}{T_R^{b}}}
{\pi R}\int_0^{\pi R} dz ~
\exp{\Big [{ -\frac {2}{\kappa}\int_z^{\pi R} M(t)dt }\Big ]}
\frac{\int_z^{\pi R} g_5^2(t)e^{A(t) }dt}{\int_0^{\pi R} g_5^2(t)e^{A(t)}dt }
\ee
Therefore  assuming $M(t)>0$, we have 
$\kappa_R^{CC}\leq T_R^{t} T_R^{b}$. 

We will make use of the 2$\sigma$ experimental bound for $\kappa_R^{CC}$, 
$\vert \kappa_R^{CC}\vert \leq 4\times 10^{-3}$ \cite{Larios:1999au}, by saturating  the strongest constraint $T_R^{t} 
T_R^{b}\sim 4\times 10^{-3}$. In this way
we determine $T_R^{t}=0.35$ and $T_R^{b}=0.01$. With this choice
the top mass value can be obtained, assuming  $R^{-1}\sim $ 1 TeV and
 $\kappa\sim 10$, for both cases. Therefore the KK fermion excitations are approximately ten times heavier than the corresponding gauge excitations, thus enforcing our derivation of the effective theory by integrating out the bulk fermions.
The value of $T_R^{b}$
is consistent with the experimental bounds on the corrections to the right-handed coupling $g_R^b$ of the bottom  to $Z$. In our model $\delta g_R^b\sim (T_R^{b})^2 \sim 10^{-4}$,
whereas the experimental bound is  $\delta g_R^b\leq 0.0219$ \cite{Haber:1999zh}. In a particular case of the BESS model considered in  \cite{Chivukula:2006cg}, similar conclusions are reached. 

From our analysis it turns out that, in the large curvature limit, the warping does not substantially modify the
discussion and the conclusions of the flat case once the global cancellation in $\eps_3^N$ is required.

\section{Conclusions}
\label{section6}

Five dimensional gauge models with flat or warped metric offers an attractive
alternative to the problem of the electroweak breaking. A general feature
of these scenarios is however, like for technicolor models, a large contribution to the $\eps_3$ parameter.
In this paper we have investigated the possibility of a cancellation between the gauge contribution and  the
fermion one in the  $\eps_3$ parameter
 once fermions are delocalized. This cancellation can be local (in the fifth dimension) or global.
 Starting from the deconstructed version of a five dimensional
$SU(2)$ gauge theory with bulk fermions in a generic warped metric, we have studied the effects of the new
physics in the low energy limit by eliminating the bulk fermions.
In this way effective couplings of the standard model fermions to the bulk bosons and also mass terms for
the fermions are generated. We have shown that
the requirement of local cancellation in $\eps_3^N$  necessarily implies
the vanishing of the fermion masses. This is due to the presence of one horizon on the right (IR) brane and it holds true in any generic warped metric.
Finally we have considered the global cancellation
both in the flat and in the warped case.


\end{document}